\documentclass[twocolumn]{jpsj2}

\title{%
Conductance Distribution in Disordered Quantum Wires\\
with a Perfectly Conducting Channel
}

\author{%
Yositake {\sc Takane},$^{1}$ Shingo {\sc Iwasaki},$^{1}$
Yuka {\sc Yoshioka},$^{1}$ Masayuki {\sc Yamamoto},$^{1}$
and Katsunori {\sc Wakabayashi}$^{1,2}$
}

\inst{%
$^{1}$Department of Quantum Matter,
Graduate School of Advanced Sciences of Matter, Hiroshima University, \\
Higashihiroshima, Hiroshima 739-8530, Japan \\
$^{2}$PRESTO, Japan Science and Technology Agency (JST), Kawaguchi,
Saitama 332-0012, Japan
}

\recdate{ \hspace{50mm} }

\abst{%
We study the conductance of phase-coherent disordered quantum wires
focusing on the case in which the number of conducting channels
is imbalanced between two propagating directions.
If the number of channels in one direction is by one greater than
that in the opposite direction,
one perfectly conducting channel without backscattering is stabilized
regardless of wire length.
Consequently, the dimensionless conductance does not vanish
but converges to unity in the long-wire limit,
indicating the absence of Anderson localization.
To observe the influence of a perfectly conducting channel,
we numerically obtain the distribution of conductance
in both cases with and without a perfectly conducting channel.
We show that the characteristic form of the distribution is notably
modified in the presence of a perfectly conducting channel.
}

\kword{%
perfectly conducting channel, DMPK equation, graphene nanoribbon,
unitary class
}

\begin{document}
\sloppy
\maketitle

\section{Introduction}

The discovery of a perfectly conducting channel in disordered carbon
nanotubes~\cite{ando1,ando2,nakanishi,ando3} and graphene
nanoribbons~\cite{wakabayashi1,wakabayashi2} provides a counterexample
to the conjecture that an ordinary quasi-one-dimensional quantum system
with disorder exhibits Anderson localization.
The band structure of carbon nanotubes and graphene nanoribbons has
two energy valleys near the Dirac points, which are well separated in
momentum space. 
Ando and co-workers have shown that a perfectly conducting channel without
backscattering appears in each valley of a metallic carbon nanotube
when impurity scattering does not connect
the two valleys.~\cite{ando1,ando2,nakanishi,ando3}
Such a situation can be realized
when only long-ranged impurities are distributed.
The presence of a perfectly conducting channel is concluded from the facts
that the reflection matrix $r$ for each valley has the skew-symmetry
(i.e., $^{t}r = - r$)
and that the number of conducting channels is odd in each valley.
It has been pointed out that a perfectly conducting channel is not intrinsic to
carbon nanotubes with long-ranged disorder, but universally appears in
disordered quasi-one-dimensional systems which belong to
the symplectic universality class with an odd number of conducting
channels.~\cite{takane1,takane2,takane3,takane4,sakai1,
ando4,sakai2,sakai3,obuse,kobayashi}
Note that the symplectic universality class consists of systems having
time-reversal symmetry without spin-rotation invariance.
The discovery of a perfectly conducting channel in the symplectic universality
class resolves the long-standing puzzle
raised by Zirnbauer and co-workers.~\cite{puzzle,zirnbauer,mirlin}
The presence of a perfectly conducting channel in graphene nanoribbons
with two zigzag edges has been pointed out by Wakabayashi and
co-workers~\cite{wakabayashi1,wakabayashi2}.
As in the case of carbon nanotubes, its presence is ensured
only when impurities are long-ranged.
However, the corresponding stabilization mechanism of a perfectly conducting
channel is completely different from that in the case of carbon nanotubes.
In each valley of a graphene nanoribbon with two zigzag edges,
the number of conducting channels in one propagating direction is
by one greater, or smaller, than that in the opposite direction
because each valley has an excess one-way channel.
This imbalance results in the appearance of
a perfectly conducting channel~\cite{barnes1,barnes2}.
The presence of a perfectly conducting channel is a universal characteristic
emerging in disordered quasi-one-dimensional systems which belong to
the unitary universality class with the channel-number imbalance
between two propagating directions.~\cite{kobayashi,barnes1,barnes2,hirose,
takane5,takane6,takane7,takane8}
The unitary universality class is characterized by
the absence of time-reversal symmetry.

The behavior of the dimensionless conductance in such systems
with a perfectly conducting channel has been studied by means of
the scaling theory based on a random-matrix
model,~\cite{takane1,takane2,sakai1,sakai2,takane5,takane7,takane8}
the super-symmetric field theory~\cite{takane3,sakai2},
the mapping to a super-spin chain model~\cite{takane6,takane7},
and numerical simulations~\cite{takane4,ando4,sakai3,obuse,kobayashi,
barnes1,barnes2,hirose}.
It has been shown that the dimensionless conductance decays exponentially
with increasing system length $L$ towards unity,
indicating the absence of Anderson localization.
This behavior should be contrasted to that in the ordinary case
without a perfectly conducting channel, in which the dimensionless conductance
vanishes in the long-$L$ limit.
It has also been shown that the dimensionless conductance
in the presence of a perfectly conducting channel decays
much faster than that in the ordinary case.
To deeply understand such peculiar transport properties of
the systems with a perfectly conducting channel, it is desirable to study
the distribution of dimensionless conductance.

For simplicity, we hereafter restrict our attention to
disordered quantum wires with unitary symmetry.
We consider wires of length $L$ with $N$ right-moving channels
and $N + m$ left-moving channels with $m = 0$, $1$.
In the ordinary case of $m = 0$, the right-moving channels and
the left-moving channels have the same set of transmission eigenvalues
$\{T_{1}, T_{2}, \dots, T_{N}\}$.
Therefore, the dimensionless conductance $G_{\rm L}$ and $G_{\rm R}$
for the left-moving and right-moving channels, respectively, are equivalent
and are given by $G_{\rm R} = G_{\rm L} = g$ with
$g = \sum_{a=1}^{N}T_{a}$.
In the case of $m = 1$, one left-moving channel becomes perfectly conducting
and $G_{\rm L}$ and $G_{\rm R}$ differ from each other.
If the set of transmission eigenvalues for the right-moving channels is
$\{T_{1}, T_{2}, \dots, T_{N}\}$, that for the left-moving channels is
expressed as $\{T_{1}, T_{2}, \dots, T_{N}, 1 \}$, where we have identified
the $N+1$th channel as the perfectly conducting one.~\cite{takane5}
We obtain $G_{\rm L} = g + 1$ and $G_{\rm R} = g$
with $g = \sum_{a=1}^{N}T_{a}$.
The above argument indicates that in both the cases
of $m = 0$ and of $m = 1$, we can completely describe the statistical property
of $G_{\rm L}$ and $G_{\rm R}$
if the distribution of $g$ is given as a function of the wire length $L$.
Let $p(g)$ be the distribution of $g$.
As the averaged conductance $\langle g \rangle$ monotonically decreases
with increasing $L$, we employ it as the scaling parameter instead of $L$.
In the absence of a perfectly conducting channel (i.e., $m = 0$),
it has been shown that
the characteristic form of $p(g)$ crucially depends on $\langle g \rangle$;
the dimensionless conductance $g$ obeys a Gaussian distribution
in the short-wire regime of $\langle g \rangle \gg 1$,~\cite{gopar}
while a log-normal distribution is expected
in the long-wire regime of $\langle g \rangle \ll 1$.~\cite{pichard}
It has also been shown that an interesting behavior appears in the crossover
regime of $\langle g \rangle \sim 1$.~\cite{muttalib,froufe-perez}
It is interesting how the behavior of $p(g)$ is affected by
the presence of a perfectly conducting channel.

In this paper, we theoretically study the behavior of $p(g)$
in disordered quantum wires with unitary symmetry
focusing on the crossover regime of $\langle g \rangle \sim 1$.
We obtain $p(g)$ in both the cases of $m = 0$ and of $m = 1$
at $\langle g \rangle = 1.0$, $0.5$ and $0.35$.
To obtain $p(g)$, we first adopt a classical Monte Carlo approach
based on the existing scaling theory.
Employing an approximate analytic expression of the probability distribution
function of transmission eigenvalues,
we calculate $p(g)$ by using a classical Monte Carlo method.
For comparison, we also calculate $p(g)$ by adopting
a numerical simulation approach with a tight-binding model.
For the case of $m = 0$, we employ the tight-binding model for a graphene
nanoribbon with zigzag and bearded edges,
while the tight-binding model for a graphene nanoribbon with
two zigzag edges is employed for the case of $m = 1$.
We confirm that the Monte Carlo approach and the numerical simulation approach
provide qualitatively identical results.
We show that at $\langle g \rangle = 1.0$ (i.e., near the diffusive regime),
the distribution $p(g)$ in the case of $m = 1$ is similar to
that in the ordinary case of $m = 0$.
However, a notable difference appears
for the smaller values of $\langle g \rangle$.
It is shown that in the case of $m = 0$,
the distribution $p(g)$ at both $\langle g \rangle = 0.5$ and $0.35$
are cut off when $g$ exceeds unity, resulting in the appearance of
a kink near $g = 1$,~\cite{froufe-perez,muttalib}
while no such structure appears in the case of $m = 1$.
This indicates that the absence of a kink in $p(g)$ is a notable
characteristic of disordered quantum wires with a perfectly conducting channel.
We explain this behavior from the viewpoint of eigenvalue repulsion
due to the perfectly conducting transmission eigenvalue.

In the next section, we obtain the conductance distribution $p(g)$
by using a classical Monte Carlo approach based on the scaling theory.
In \S 3, we obtain $p(g)$ by using a numerical simulation
approach with a tight-binding model.
Section 4 is devoted to summary.

\section{Monte Carlo Approach}

We present the scaling theory based on a random-matrix model
to describe electron transport properties in systems with
the channel-number imbalance.~\cite{takane5}
Let us consider disordered wires with $N$ right-moving channels
and $N + m$ left-moving channels with $m = 0$, $1$.
As discussed in the previous section, the dimensionless conductance
$G_{\rm L}$ and $G_{\rm R}$ are expressed in terms of
$g \equiv \sum_{a=1}^{N}T_{a}$ as $G_{\rm L} = m + g$ and $G_{\rm R} = g$.
It should be noted that the mean free paths $l$ and $l'$ for the left-moving
and right-moving channels, respectively, are not equal if $m \neq 0$.
Indeed, they satisfy $l'= (N/(N+m))l$.
The statistical behavior of $g$ is described by
the probability distribution of $\{T_{1}, T_{2}, \dots, T_{N}\}$.
We define $\lambda_{a} \equiv (1-T_{a})/T_{a}$
and introduce the probability distribution $P(\{\lambda_{a}\};s)$
of $\{\lambda_{1}, \lambda_{2}, \dots, \lambda_{N}\}$,
where $s$ is the normalized system length defined by $s \equiv L/l$.
The Fokker-Planck equation for $P(\{\lambda_{a}\};s)$, which is called
the Dorokhov-Mello-Pereyra-Kumar (DMPK) equation,~\cite{dorokhov,mello}
is expressed as~\cite{takane5}
\begin{align}
      \label{eq:dmpk-lambda}
 & \frac{\partial P(\{\lambda_{a}\};s)}{\partial s}
         \nonumber \\
 & \hspace{0mm}
   = \frac{1}{N}\sum_{a=1}^{N} \frac{\partial}{\partial \lambda_{a}}
   \left( \lambda_{a}(1+\lambda_{a}) J
          \frac{\partial}{\partial \lambda_{a}}
            \left( \frac{P(\{\lambda_{a}\};s)}{J} \right)
   \right) 
\end{align}
with
\begin{equation}
      \label{eq:jacob}
  J = \prod_{c=1}^{N} \lambda_{c}^{m}
      \times
      \prod_{b=1}^{N-1}\prod_{a=b+1}^{N}|\lambda_{a}-\lambda_{b}|^{2} .
\end{equation}
The factor $\prod_{c=1}^{N} \lambda_{c}^{m}$ in $J$ represents the repulsion
arising from the perfectly conducting eigenvalue.
This eigenvalue repulsion suppresses non-perfectly conducting eigenvalues
$\{T_{1}, T_{2}, \dots, T_{N}\}$.
It should be emphasized that the influence of a perfectly conducting
channel is described by only this factor.

We numerically calculate $p(g)$ by using a classical Monte Carlo approach
based on an approximate probability distribution of transmission eigenvalues.
The probability distribution can be obtained from
the exact solution of the DMPK equation \cite{beenakker,akuzawa}.
We define $x_{a}$ by the relation of $\lambda_{a} \equiv \sinh^{2}x_{a}$.
In terms of a set of variables $\{ x_{a} \}$,
the probability distribution is given by \cite{takane8}
\begin{equation}
      \label{eq:prob_dis_mod}
   P(\{x_{a}\};s)
 = {\rm const.} \, {\rm e}^{ - H(\{x_{a}\})}
\end{equation}
with
\begin{align}
      \label{eq:Hamiltonian}
  H(\{x_{a}\})
   & = \sum_{a=1}^{N}
       \left(   \frac{N}{s} x_{a}^{2}
              - \left(m+\frac{1}{2}\right)
                \ln \left|x_{a}\sinh 2x_{a} \right|
       \right)
          \nonumber \\
   & \hspace{-10mm}
     - \sum_{a,b=1(a>b)}^{N}
       \Bigl(   \ln \left|\sinh^{2}x_{a}-\sinh^{2}x_{b}\right|
              + \ln \left|x_{a}^{2}-x_{b}^{2}\right|
       \Bigr) .
\end{align}
Although eq.~(\ref{eq:prob_dis_mod}) is derived
under the assumption of $s/4N \ll 1$, we can expect that
it is qualitatively reliable even when $s/4N \sim 1$.~\cite{takane8}
Note that $g$ is expressed as $g = \sum_{a=1}^{N} 1/\cosh^{2}x_{a}$.

We can interpret $H(\{x_{a}\})$ as the Hamiltonian function of
$N$ classical particles in one dimension.
This analogy allows us to adapt a Monte Carlo approach
to numerical calculations of $p(g)$.
Using a simple Metropolis algorithm,~\cite{froufe-perez,canali}
we obtain $p(g)$ for $N = 5$ in both the cases of $m = 0$ and of $m = 1$
at $\langle g \rangle = 1.0$, $0.5$ and $0.35$.
By numerically calculating $\langle g \rangle$ as a function of
the normalized system length $s$, we find that
$\langle g \rangle = 1.0$ at $s/N = 0.77$ ($0.45$),
$\langle g \rangle = 0.5$ at $s/N = 1.59$ ($0.75$)
and $\langle g \rangle = 0.35$ at $s/N = 2.15$ ($0.91$)
for $m = 0$ ($m = 1$).
As pointed out in ref~\citen{takane5}, the decrease of $\langle g \rangle$
as a function of $s$ becomes fast with increasing $m$ due to the eigenvalue
repulsion arising from the perfectly conducting eigenvalue $T = 1$.
Therefore, the value of $s/N$ yielding a given $\langle g \rangle$
for $m = 1$ is smaller than that for $m = 0$.

\begin{figure}[h]
\begin{center}
\includegraphics[width=5.0cm]{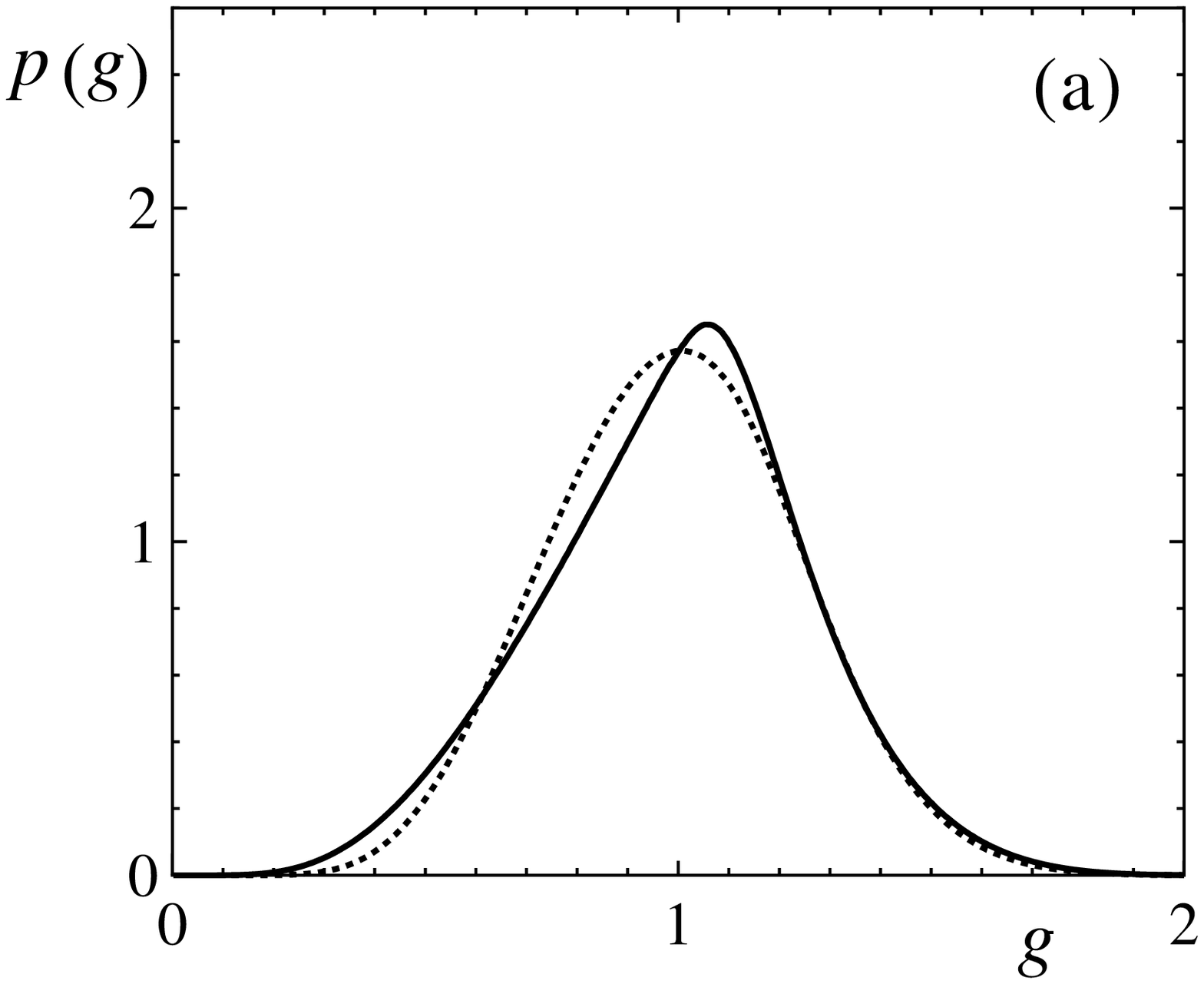}
\includegraphics[width=5.0cm]{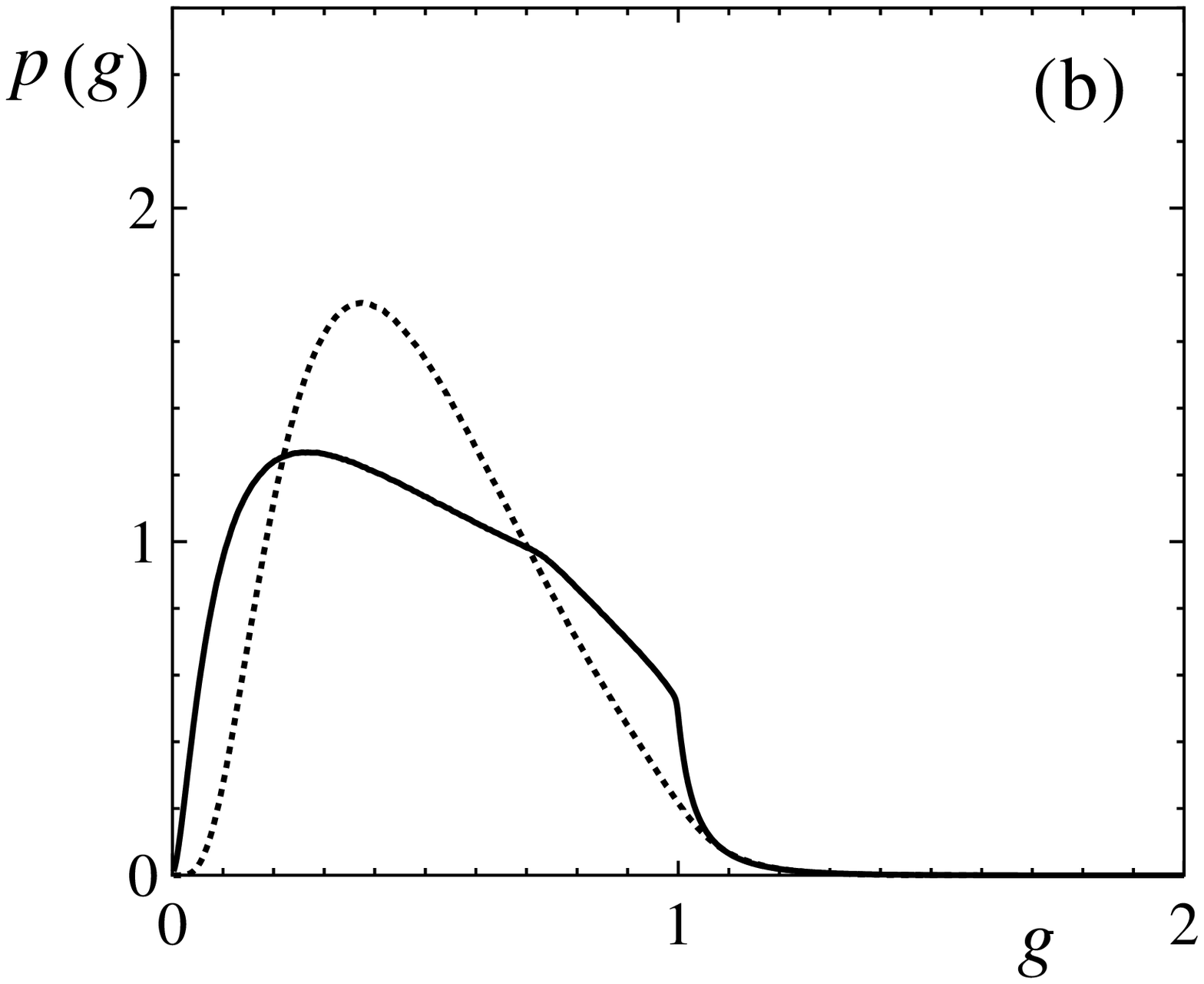}
\includegraphics[width=5.0cm]{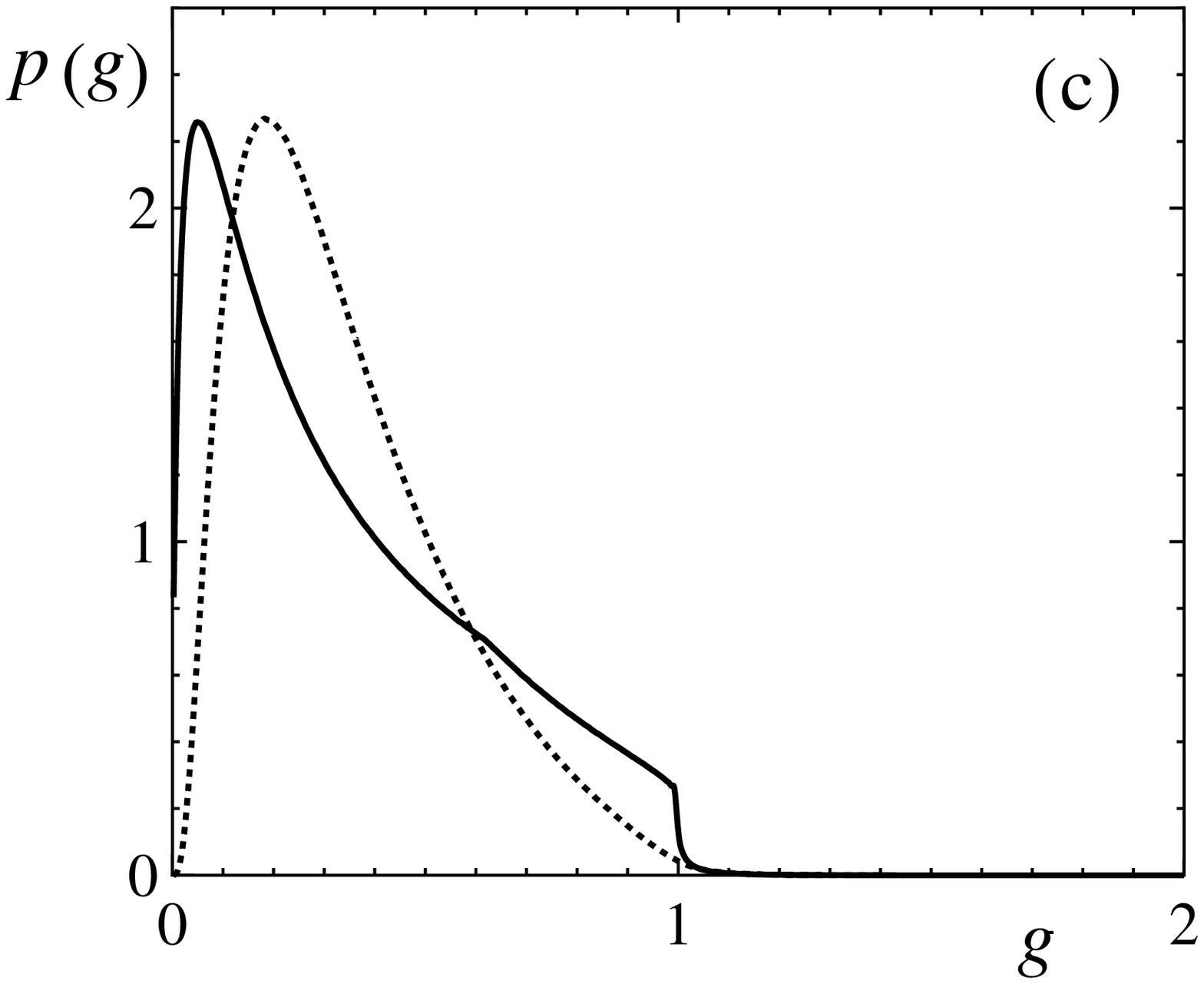}
\caption{
The conductance distributions
at (a) $\langle g \rangle = 1.0$, (b) $\langle g \rangle = 0.5$
and (c) $\langle g \rangle = 0.35$ for $m = 0$ (solid lines) and $m = 1$
(dotted lines).}
\end{center}
\end{figure}
The resulting conductance distributions are shown in Fig.~1.
Each distribution is obtained by using $2 \times 10^{9}$ Monte Carlo steps.
Except for the case of $\langle g \rangle = 1.0$, a notable difference
appears between the result for $m = 0$ and that for $m = 1$.
In the ordinary case of $m = 0$, the distributions at both
$\langle g \rangle = 0.5$ and $0.35$ are cut off when $g$ exceeds unity,
resulting in the appearance of a kink near
$g = 1$~\cite{froufe-perez,muttalib}.
We observe that no such structure appears when $m = 1$.
The behavior similar to this has been observed in the numerical result
of refs.~\citen{kobayashi} and \citen{hirose}.

We explain this behavior as follows.
In the crossover regime, $g$ is dominated by the largest transmission
eigenvalue $T_{1}$ and the second-largest one $T_{2}$.
That is, $g \approx T_{1} + T_{2}$.
In the case of $m = 0$, the largest eigenvalue $T_{1}$
can become nearly equal to unity.
However, the second-largest eigenvalue $T_{2}$ is suppressed by
the eigenvalue repulsion from $T_{1}$ and thus $T_{2} \ll T_{1} \le 1$.
Therefore, $g$ rarely becomes larger than unity,
resulting in the appearance of the kink near $g = 1$.
In contrast, $T_{1}$ in the case of $m = 1$ cannot become nearly equal to
unity due to the eigenvalue repulsion from the perfectly conducting eigenvalue.
Therefore, $p(g)$ near $g = 1$ is suppressed and the kink disappears.
We also observe that in the case of $m = 1$, the distribution is suppressed
near $g = 0$, as well as $g = 1$, and the width of $p(g)$ becomes narrow.
The suppression near $g = 0$ simply reflects the fact that
$s/N$ yielding a given $\langle g \rangle$ is smaller in the case of $m = 1$
than in the case of $m = 0$.
From eq.~(\ref{eq:Hamiltonian}), we understand that smaller $s/N$ results in
smaller $\{ x_{a} \}$ and therefore $g$ tends to become large.
Thus, $p(g)$ near $g = 0$ is more suppressed in the case of $m = 1$
than in the case of $m = 0$.

We finally note that in the case of $m = 0$
without a perfectly conducting channel,
$p(g)$ at $\langle g \rangle = 0.5$ is similar to the critical conductance
distribution for the quantum Hall transition~\cite{ohtuski} and
$p(g)$ at $\langle g \rangle = 0.35$ is similar to
the critical conductance distribution
for the Anderson transition in three dimensions.~\cite{slevin}

\section{Numerical Simulation Approach}

In this section, we calculate conductance distributions by adopting
a numerical simulation approach.
As model systems, we consider a graphene nanoribbon with zigzag and bearded
edges (zigzag-bearded nanoribbon) and a graphene nanoribbon
with two zigzag edges (zigzag-zigzag nanoribbon) (see Fig.~2).
\begin{figure}[tbh]
\begin{center}
\includegraphics[width=4.0cm]{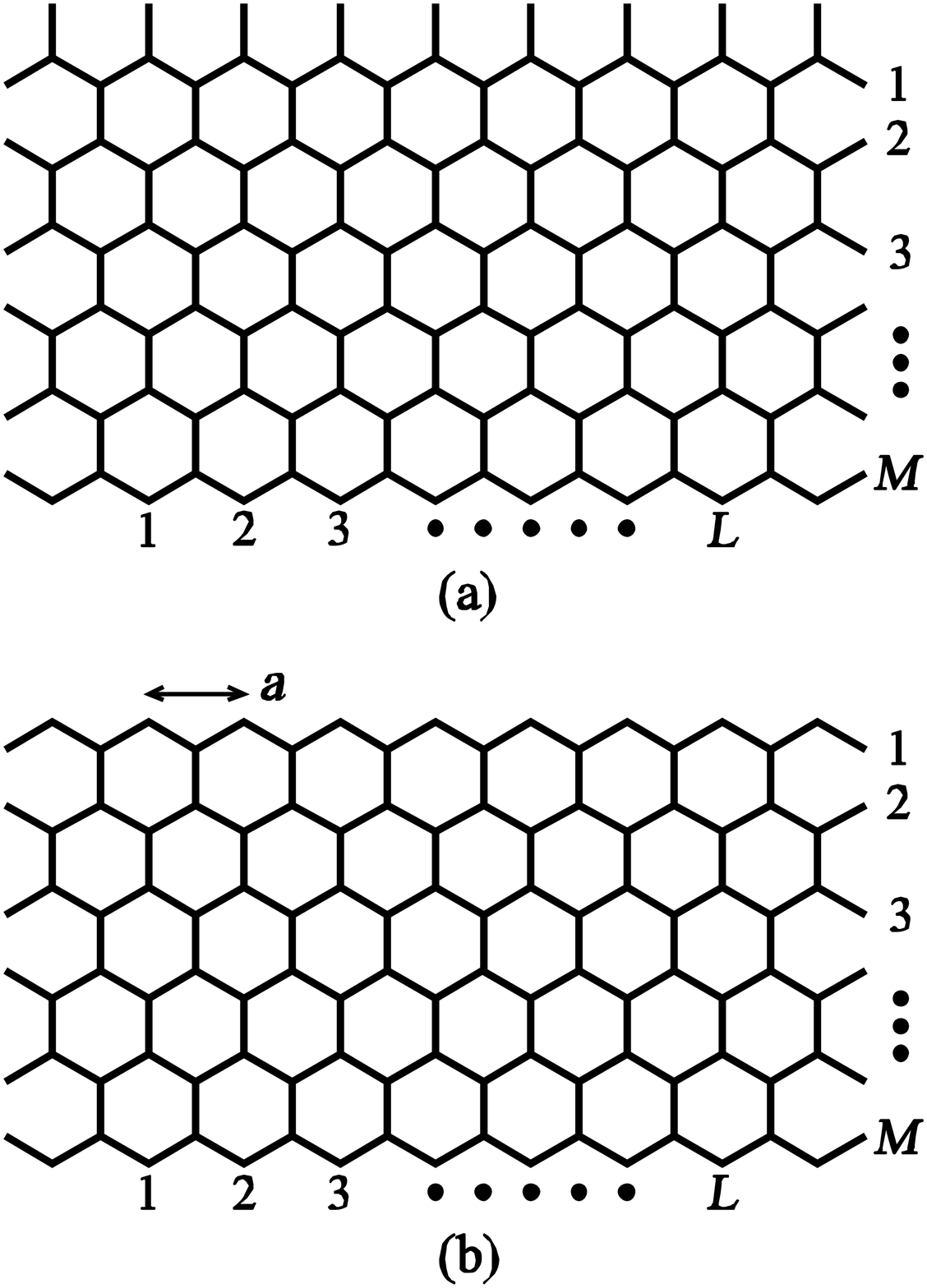}
\caption{
The structures of (a) zigzag-bearded nanoribbon
and (b) zigzag-zigzag nanoribbon.}
\end{center}
\end{figure}
In the following, we fix the number of lattice sites in the transverse
direction to be $M = 30$.
We describe electronic states of these nanoribbons by using
a tight-binding model on the honeycomb lattice.
For the case of $m = 0$,
we employ the tight-binding model for the zigzag-bearded nanoribbon,
while that for the zigzag-zigzag nanoribbon is employed
for the case of $m = 1$.
The tight-binding model is given by
\begin{align}
  H = - \sum_{i,j} \gamma_{i,j}|i\rangle \langle j|
      + \sum_{i}V_{i}|i\rangle \langle i| ,
\end{align}
where $|i\rangle$ and $V_{i}$ represent the localized electron state
and the impurity potential, respectively, on site $i$,
and $\gamma_{i,j}$ is the transfer integral between sites $i$ and $j$.
We assume that $\gamma_{i,j} = t$ if $i$ and $j$ are nearest neighbors
and $\gamma_{i,j} = 0$ otherwise.
The band structures obtained from this tight-binding
model without disorder are displayed in Fig.~3.
\begin{figure}[b]
\begin{center}
\includegraphics[width=5.0cm]{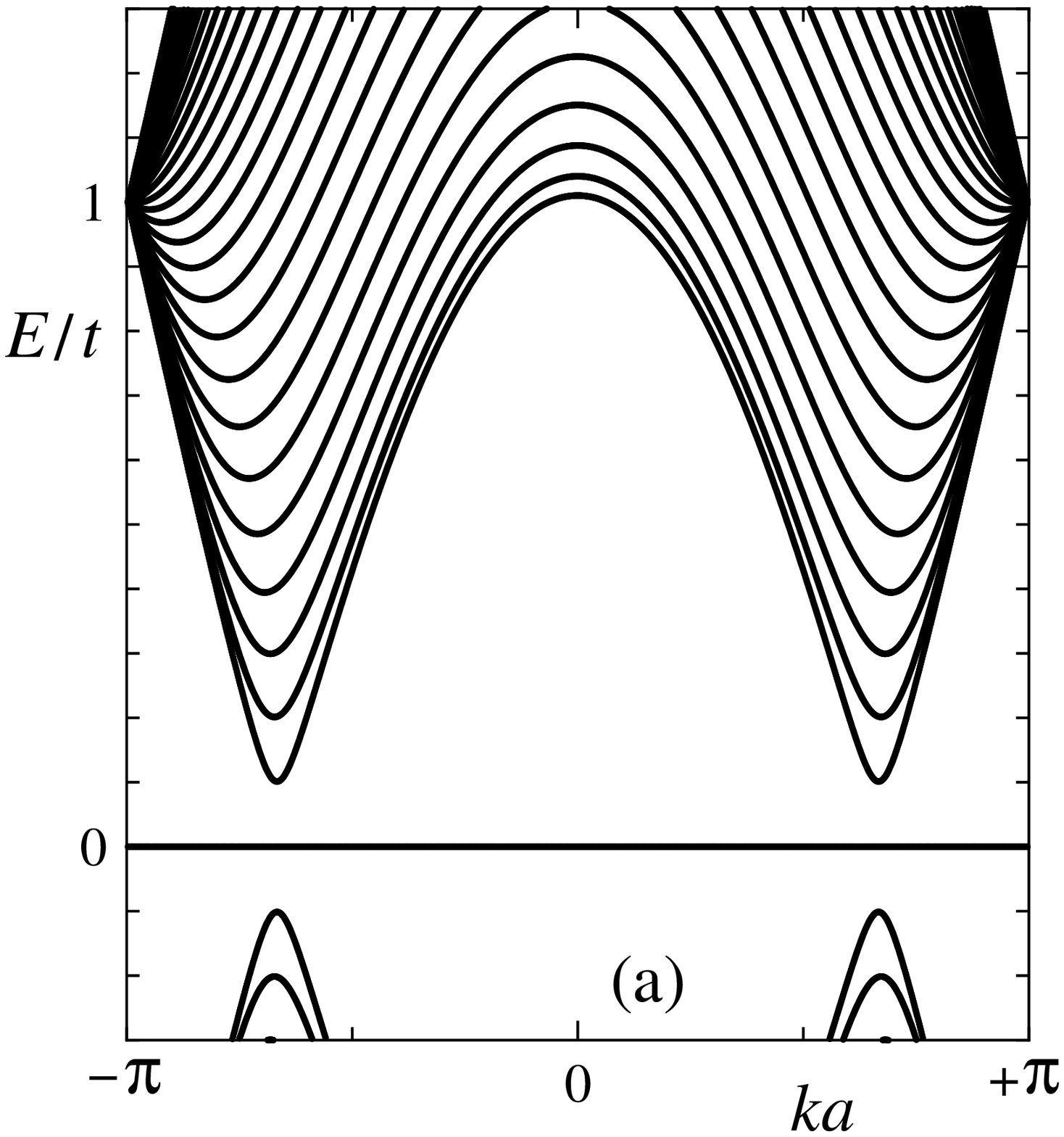}
\includegraphics[width=5.0cm]{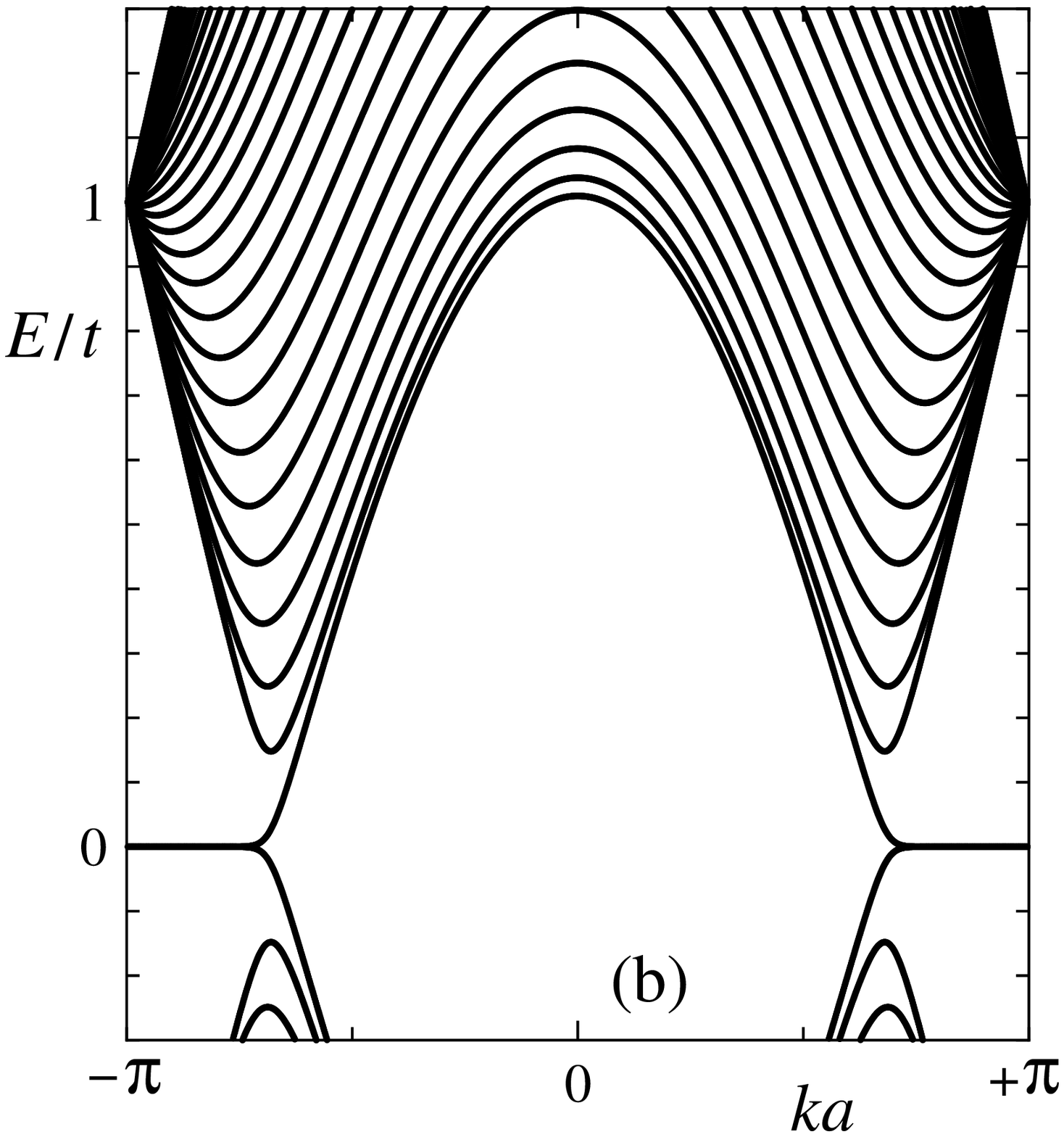}
\caption{
The band structures of (a) zigzag-bearded nanoribbon
and (b) zigzag-zigzag nanoribbon with $M = 30$.}
\end{center}
\end{figure}
We observe that the zigzag-bearded nanoribbon has a completely flat
band,~\cite{wakabayashi2,wakabayashi3}
while a partially flat band appears in
the zigzag-zigzag nanoribbon,~\cite{fujita}
and that both of them commonly possess two energy valleys.
The right and left valleys are referred to
as $K_{+}$ and $K_{-}$ valleys, respectively.
We assume that impurity potential arising from each scatterer
has a spatial range greater than the lattice constant $a$.
In this case, inter-valley scattering is not induced by such
long-ranged impurity potential, and the two energy valleys are disconnected.
Thus, these energy valleys independently contribute electron transport.
Hereafter, we restrict our consideration to the case
in which the Fermi energy $E$ is placed within $0 < E/t < 1$.
Note that in the zigzag-zigzag nanoribbon, the partially flat band provides
an excess left-moving (right-moving) channel in the $K_{+}$ ($K_{-}$) valley.
Therefore, the number $N_{\rm L}$ of left-moving channels is by one
greater (smaller) than the number $N_{\rm R}$ of right-moving channels
in the $K_{+}$ ($K_{-}$) valley.
That is, $N_{\rm L} = N_{\rm R} + 1$ in the $K_{+}$ valley
and $N_{\rm L} + 1 = N_{\rm R}$ in the $K_{-}$ valley.
This results in the appearance of one perfectly conducting channel
in both the left-moving channels in the $K_{+}$ valley
and the right-moving channels
in the $K_{-}$ valley.~\cite{wakabayashi1,wakabayashi2}
In contrast, $N_{\rm L}$ and $N_{\rm R}$ are equal to each other
in both the valleys of the zigzag-bearded nanoribbon
and no perfectly conducting channel appears.
Let $G_{\rm L}^{\pm}$ and $G_{\rm R}^{\pm}$ be the dimensionless
conductance for the left-moving channels and that for the right-moving channels
in the valley $K_{\pm}$.
For the case of the zigzag-bearded nanoribbon
with $N_{\rm L} = N_{\rm R} = N$ in both the valleys, we observe that
$G_{\rm L}^{+} = G_{\rm R}^{+} = G_{\rm L}^{-} = G_{\rm R}^{-} = g$
with $g = \sum_{a = 1}^{N} T_{a}$.
In contrast, for the case of the zigzag-zigzag nanoribbon with
$N_{\rm L} - 1 = N_{\rm R} = N$ in the $K_{+}$ valley and
$N_{\rm L} = N_{\rm R} - 1 = N$ in the $K_{-}$ valley (see Table I),
we observe that $G_{\rm R}^{+} = G_{\rm L}^{-} = g$ and
$G_{\rm L}^{+} = G_{\rm R}^{-} = 1 + g$ with $g = \sum_{a = 1}^{N} T_{a}$.
Note that $G_{\rm L}^{+}$ and $G_{\rm R}^{-}$ contains the contribution
from one perfectly conducting channel.
The total conductance of these systems is expressed as
$G_{\rm total} = m + 2g$
with $m = 0$ for the zigzag-bearded nanoribbon
and $m = 1$ for the zigzag-zigzag nanoribbon.
\begin{table}[ht]
\caption{The number of left-moving channels, $N_{\rm L}$,
and the number of right-moving channels, $N_{\rm R}$,
where $m = 0$ for the zigzag-bearded
nanoribbon and $m = 1$ for the zigzag-zigzag nanoribbon.
}
\begin{center}
\begin{tabular}{c|c|c}
\hline
Valley & $N_{\rm L}$ & $N_{\rm R}$ \\
\hline
$K_{+}$ & $N+m$ & $N$ \\
$K_{-}$ & $N$ & $N+m$ \\
\hline
\end{tabular}
\end{center}
\label{t1}
\end{table}

Let us consider nanoribbons which is infinitely long
in the longitudinal direction.
To introduce disorder, we randomly distribute scatterers
in the finite region of $L$ lattice sites.
We identify $L$ with the length of our system.
We assume that the impurity potential on site $i$ arising from a scatterer
at site $j$ is given by
\begin{align}
 V_{i}(j) = w_{j}\exp\left( - |\mib{r}_{i}-\mib{r}_{j}|^{2}/d^{2} \right) ,
\end{align}
where $\mib{r}_{i}$ is the position vector of site $i$,
and $w_{j}$ and $d$ represent the amplitude and the spatial range of
this potential, respectively.
Thus, the impurity potential on site $i$ is given by
\begin{align}
 V_{i} = \sum_{j} V_{i}(j) ,
\end{align}
where $j$ is summed over the disordered region of length $L$.
The amplitude $w_{j}$ is distributed uniformly within the range of
$-W/2 < w_{j} < W/2$.
Let $p$ be the probability that each site is occupied by a scatterer.
The strength of disorder is controlled by $W$ and $p$.
Adopting the model described above, we calculate the dimensionless
conductance $G_{\rm total}$ by using the Landauer formula
\begin{align}
  G_{\rm total} = {\rm tr}\{ \mib{t} \mib{t}^{\dagger}\} ,
\end{align}
where $\mib{t}$ is the transmission matrix through the disordered region.
The dimensions of $\mib{t}$ are equal to $(2N+m) \times (2N+m)$.
We consider the behavior of $g \equiv ( G_{\rm total} - m )/2$
in the case of $N = 5$.
With $M = 30$, this is realized when $0.486 \le E/t \le 0.571$
for the zigzag-bearded nanoribbon ($m = 0$) and $0.529 \le E/t \le 0.611$
for the zigzag-zigzag nanoribbon ($m = 1$).
The following parameters are employed: $W/t = 0.1$, $p = 0.1$ and $d/a = 1.5$
with $a$ being the lattice constant (see Fig~2(b)).
We can numerically obtain $\mib{t}$ for a given impurity configuration
by using a recursive Green's function technique.

\begin{figure}[h]
\begin{center}
\includegraphics[width=5.0cm]{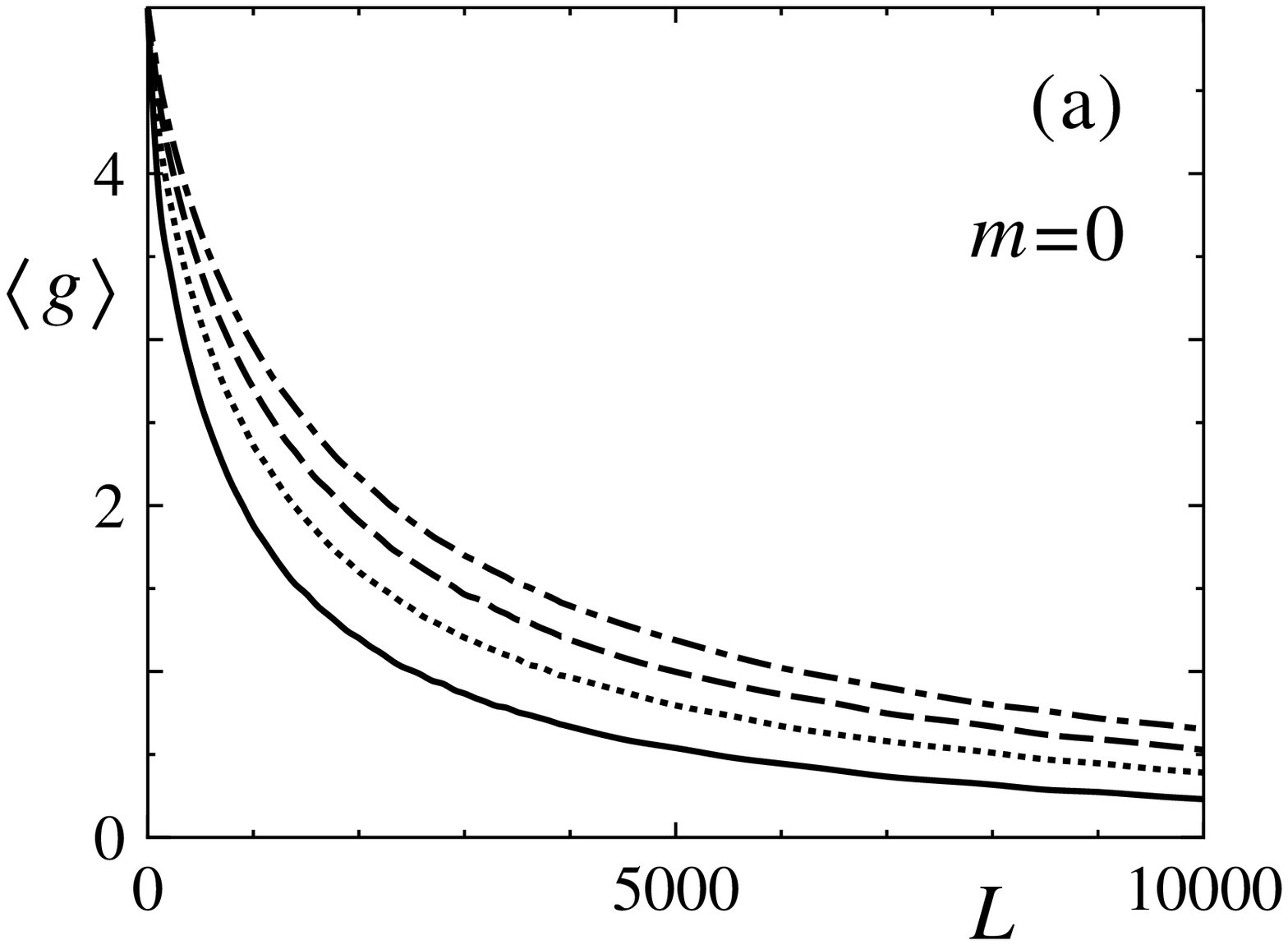}
\includegraphics[width=5.0cm]{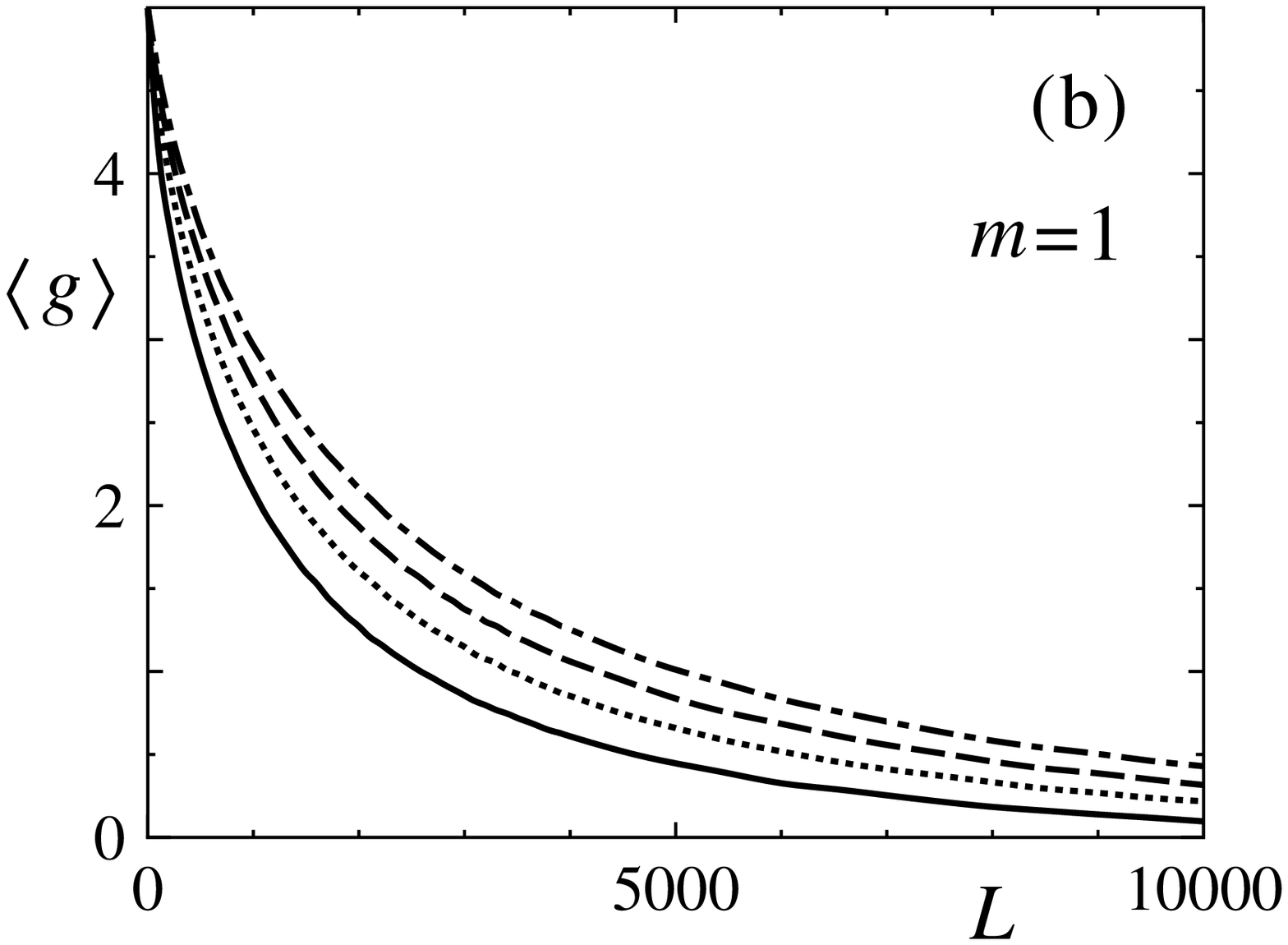}
\end{center}
\caption{
The average of $g$ for $N = 5$ with (a) $m = 0$ and (b) $m = 1$
as a function of the length $L$,
where $E/t = 0.49$, $0.50$, $0.51$ and $0.52$ from bottom to top
in the case of $m = 0$
and $E/t = 0.54$, $0.55$, $0.56$ and $0.57$ from bottom to top
in the case of $m = 1$.
}
\end{figure}
\begin{figure}[h]
\begin{center}
\includegraphics[width=5.0cm]{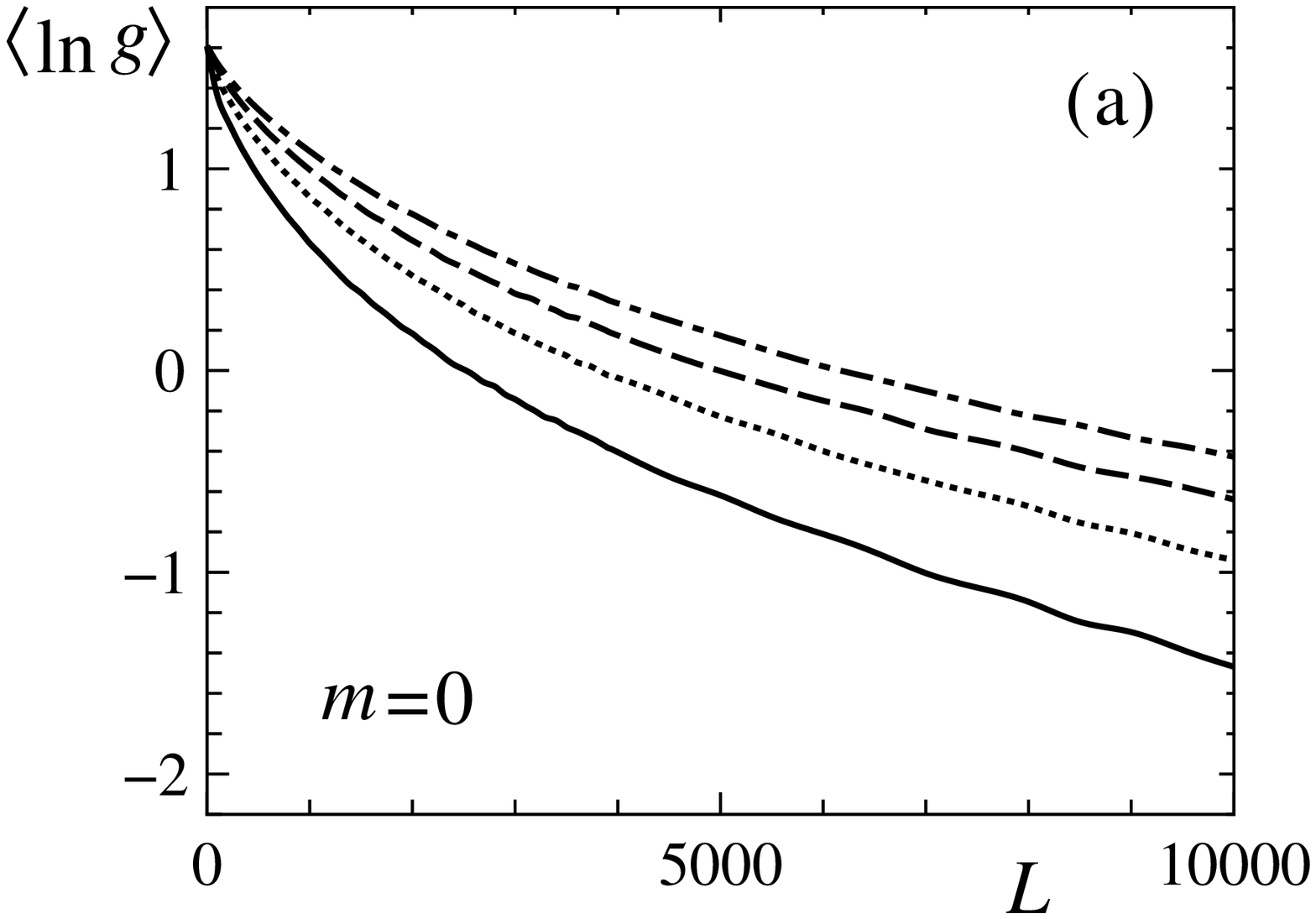}

\includegraphics[width=5.0cm]{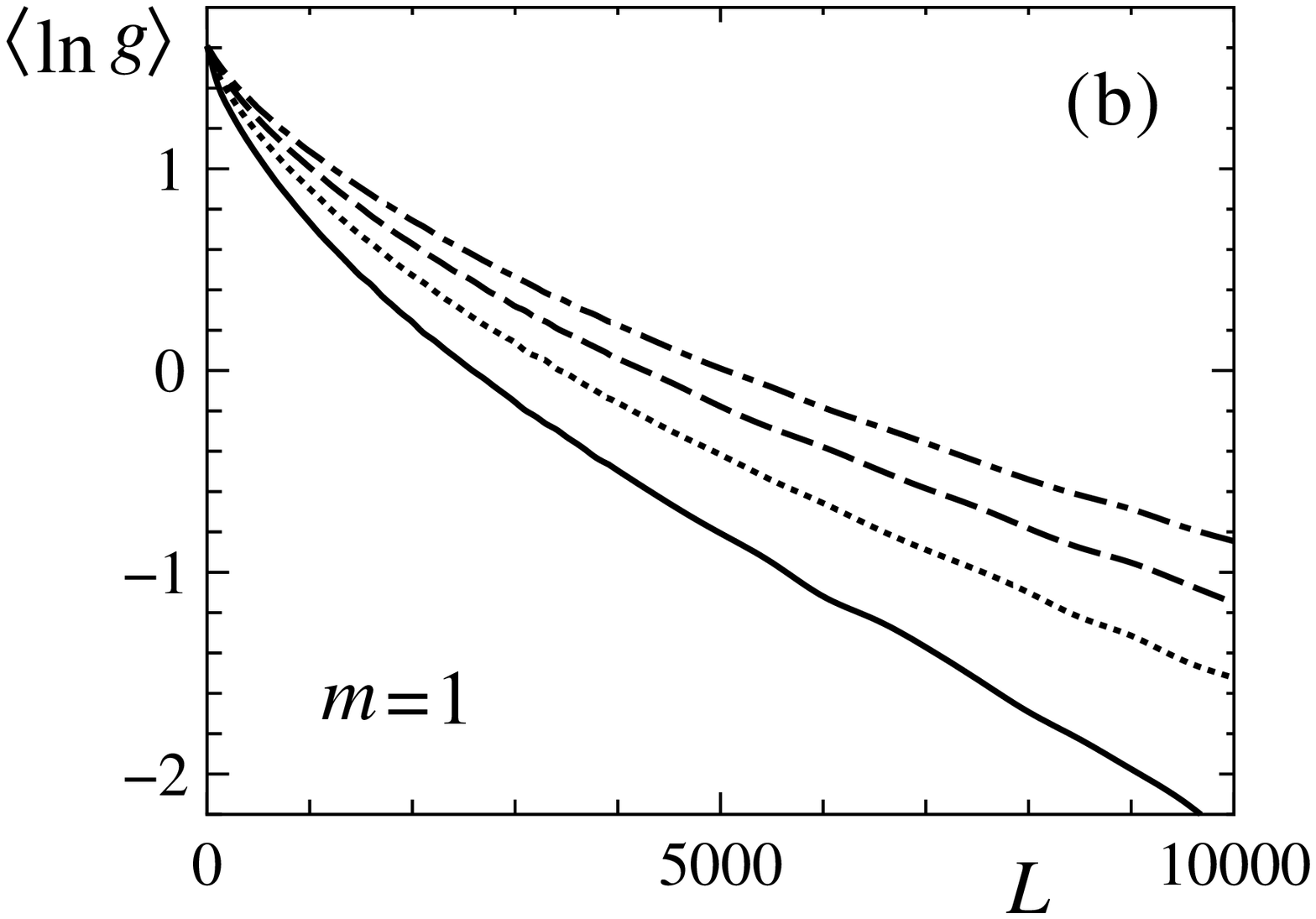}
\end{center}
\caption{
The average of $\ln g$ for $N = 5$ with (a) $m = 0$ and (b) $m = 1$
as a function of the length $L$,
where $E/t = 0.49$, $0.50$, $0.51$ and $0.52$ from bottom to top
in the case of $m = 0$
and $E/t = 0.54$, $0.55$, $0.56$ and $0.57$ from bottom to top
in the case of $m = 1$.
}
\end{figure}
Before studying the conductance distribution,
we consider the behavior of the ensemble averages of $g$ and $\ln g$.
We numerically calculate $\langle g \rangle$ and $\langle \ln g \rangle$
as a function of $L$ at $E/t = 0.49$, $0.50$, $0.51$ and $0.52$
for the case of $m = 0$ and at $E/t = 0.54$, $0.55$, $0.56$ and $0.57$
for the case of $m = 1$.
Figures~4 and 5 show the $L$-dependence of
$\langle g \rangle$ and $\langle \ln g \rangle$, respectively.
The ensemble average for each data point is performed over $5000$
samples with different impurity configuration.
We observe from Fig.~4 that the decay of $\langle g \rangle$
with increasing $L$ becomes slower with increasing $E/t$
in both the cases of $m = 0$ and of $m = 1$.
This implies that the mean free path $l$ becomes longer with increasing $E/t$.
Although the decay of $\langle g \rangle$ seems to be slightly faster
in the case of $m = 1$ than in the ordinary case of $m = 0$,
we cannot state this definitively.
However, Fig.~5 clearly shows that $\langle \ln g \rangle$ decreases faster
in the case of $m = 1$ than in the case of $m = 0$.
This is consistent with the scaling theory~\cite{takane5}
and is explained as follows.
In the case of $m = 1$, one perfectly conducting channel is present
and the corresponding transmission eigenvalue (i.e., $T = 1$) suppresses
the other transmission eigenvalues
$\{T_{1},T_{2},\dots , T_{N} \}$ due to eigenvalue repulsion.
This results in the suppression of $g$.
Indeed, the scaling theory predicts that the conductance decay length $\xi$
defined by $\exp [\langle \ln g \rangle] \equiv \exp [- 2L/\xi]$
is given by~\cite{takane5}
\begin{align}
    \label{eq:xi_rmt}
  \xi = \frac{2Nl}{m+1} ,
\end{align}
indicating that $\xi$ in the case of $m = 1$ is twice shorter than
that in the case of $m = 0$ if the mean free path is common to the two cases.
We estimate $\xi$ from the result of $\langle \ln g \rangle$.
The estimated values of $\xi$ are listed in Table~\ref{t2}.
Because the mean free path should depend on $E/t$ and $m$,
a detailed comparison with eq.~(\ref{eq:xi_rmt}) is not straightforward.
However, we observe the obvious tendency that $\xi$ in the case of $m = 1$ is
shorter than that in the case of $m = 0$.
This is consistent with the prediction of the scaling theory.
\begin{table}[ht]
\caption{
The conductance decay length $\xi$ in the cases of $m = 0$ and of  $m = 1$
for several values of $E/t$.
}
\begin{center}
\begin{tabular}{c|c||c|c}
\hline
\multicolumn{2}{c||}{$m = 0$} & \multicolumn{2}{c}{$m = 1$} \\
\hline
$E/t$ & $\xi$ & $E/t$ & $\xi$ \\
\hline
$0.49$ & $11500$ & $0.54$ & $6400$ \\
$0.50$ & $13500$ & $0.55$ & $8400$ \\
$0.51$ & $15100$ & $0.56$ & $9500$ \\
$0.52$ & $15900$ & $0.57$ & $10600$ \\
\hline
\end{tabular}
\end{center}
\label{t2}
\end{table}

Now, we consider the conductance distribution $p(g)$.
We numerically obtain $p(g)$ at $\langle g \rangle = 1.0$, $0.5$ and $0.35$
in both the cases of $m = 0$ and of $m = 1$.
We set $E/t = 0.49$ for $m = 0$ and $E/t = 0.54$ for $m = 1$.
From the numerical result displayed in Fig.~4,
we find that $\langle g \rangle = 1.0$ at $L = 2511$ ($2583$),
$\langle g \rangle = 0.5$ at $L = 5352$ ($4623$)
and $\langle g \rangle = 0.35$ at $L = 7312$ ($5801$)
for $m = 0$ ($m = 1$).
The resulting conductance distributions for the case of $m = 0$ and
those for the case of $m = 1$ are displayed in Figs.~6 and 7, respectively,
where each distribution is obtained by using $5 \times 10^{5}$ samples.
For comparison, we also display the conductance distributions obtained
by the Monte Carlo approach.
\begin{figure}[h]
\begin{center}
\includegraphics[width=5.0cm]{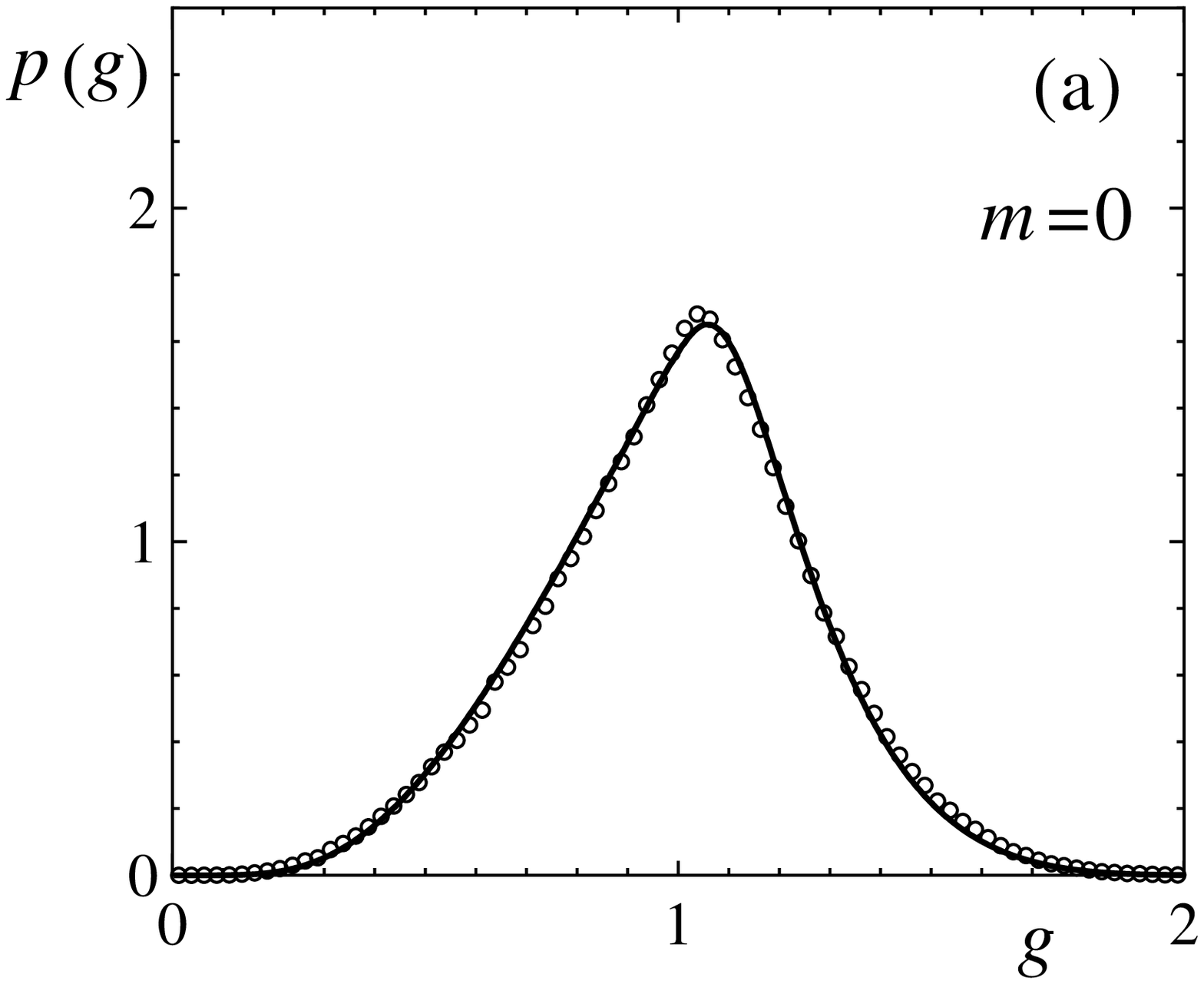}
\includegraphics[width=5.0cm]{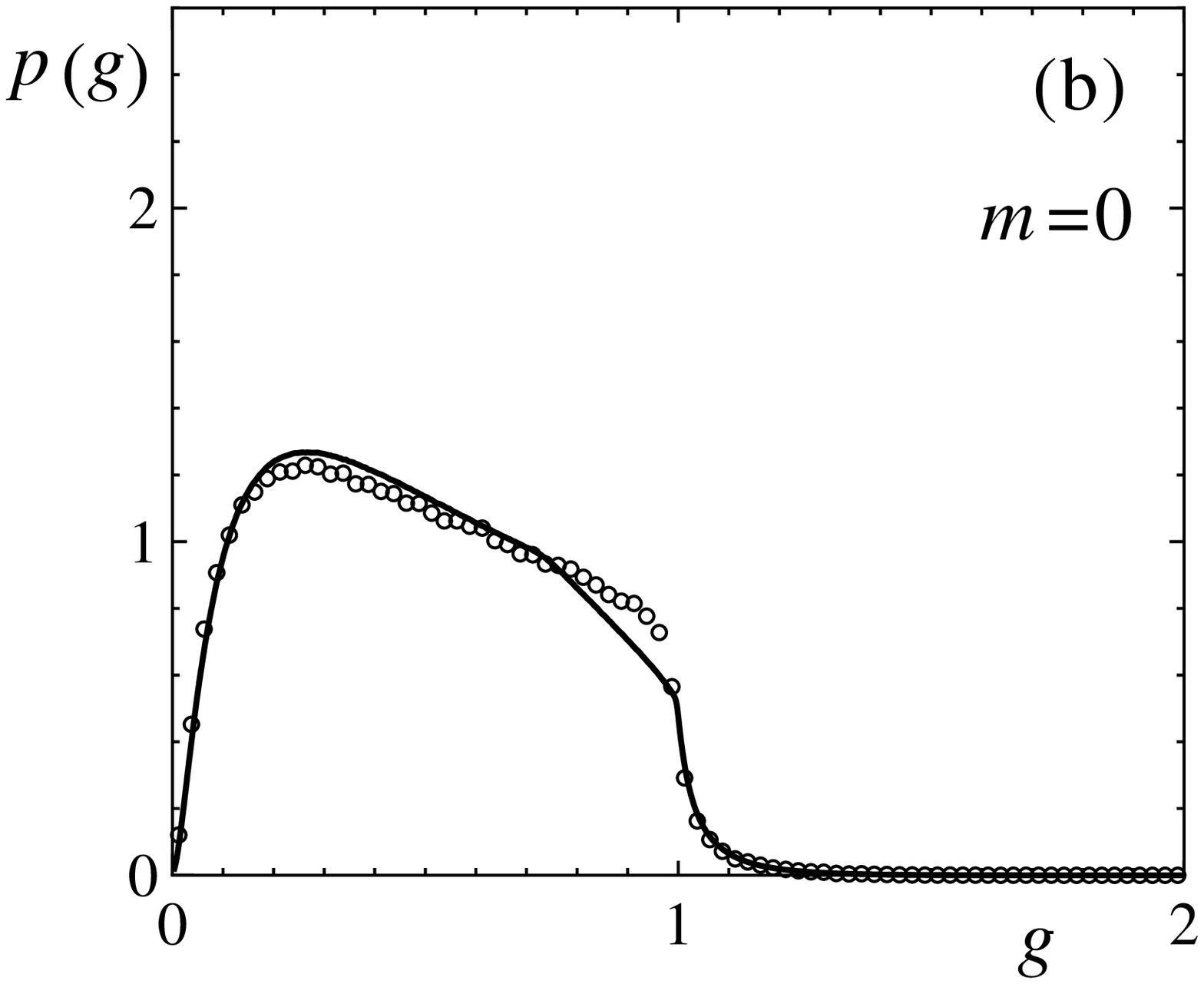}
\includegraphics[width=5.0cm]{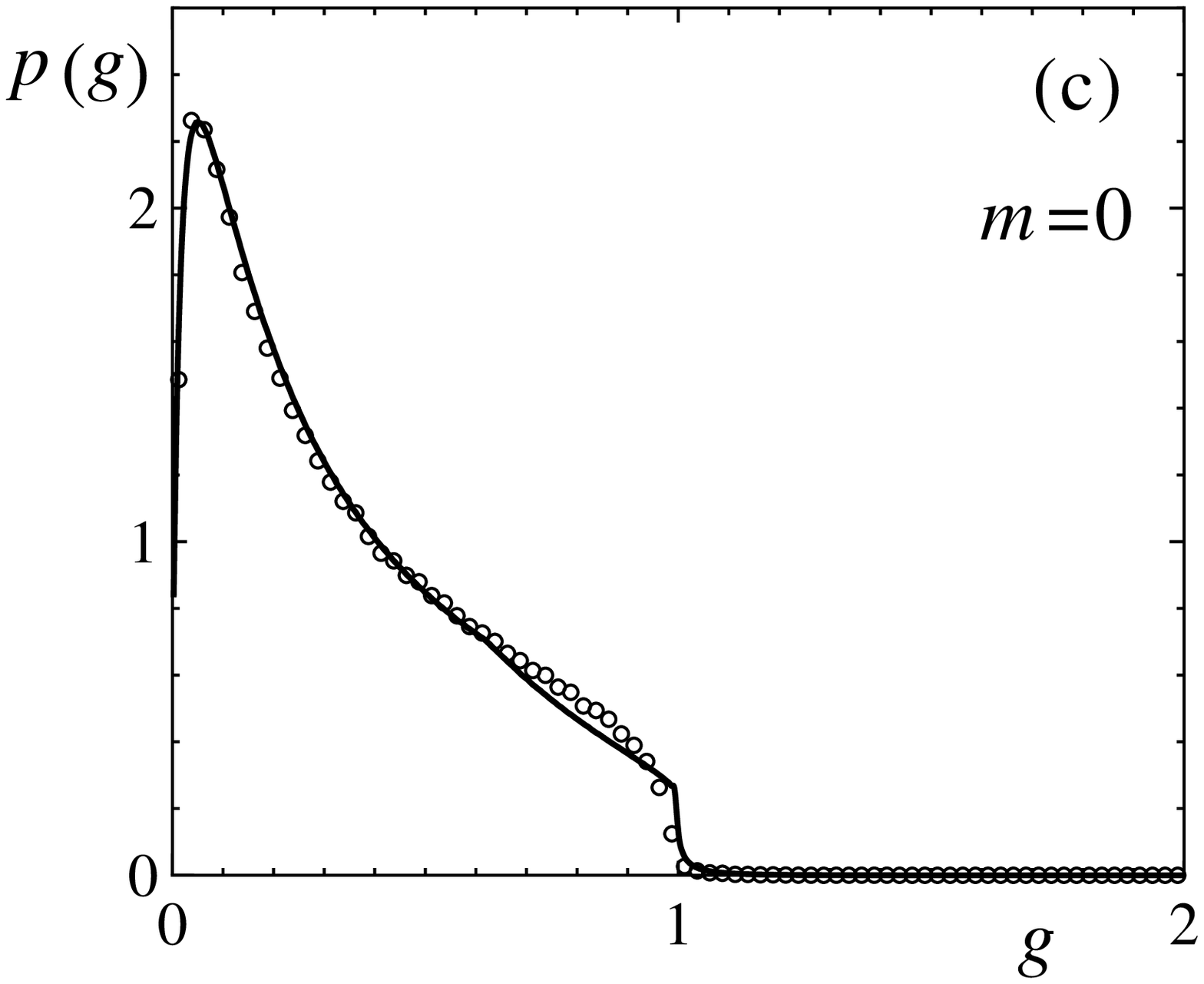}
\caption{
The conductance distributions for the case of $m = 0$
at (a) $\langle g \rangle = 1.0$, (b) $\langle g \rangle = 0.5$
and (c) $\langle g \rangle = 0.35$.
Open circles represent the numerical simulation result,
while solid lines represent the Monte Carlo result.
}
\end{center}
\end{figure}
\begin{figure}[h]
\begin{center}
\includegraphics[width=5.0cm]{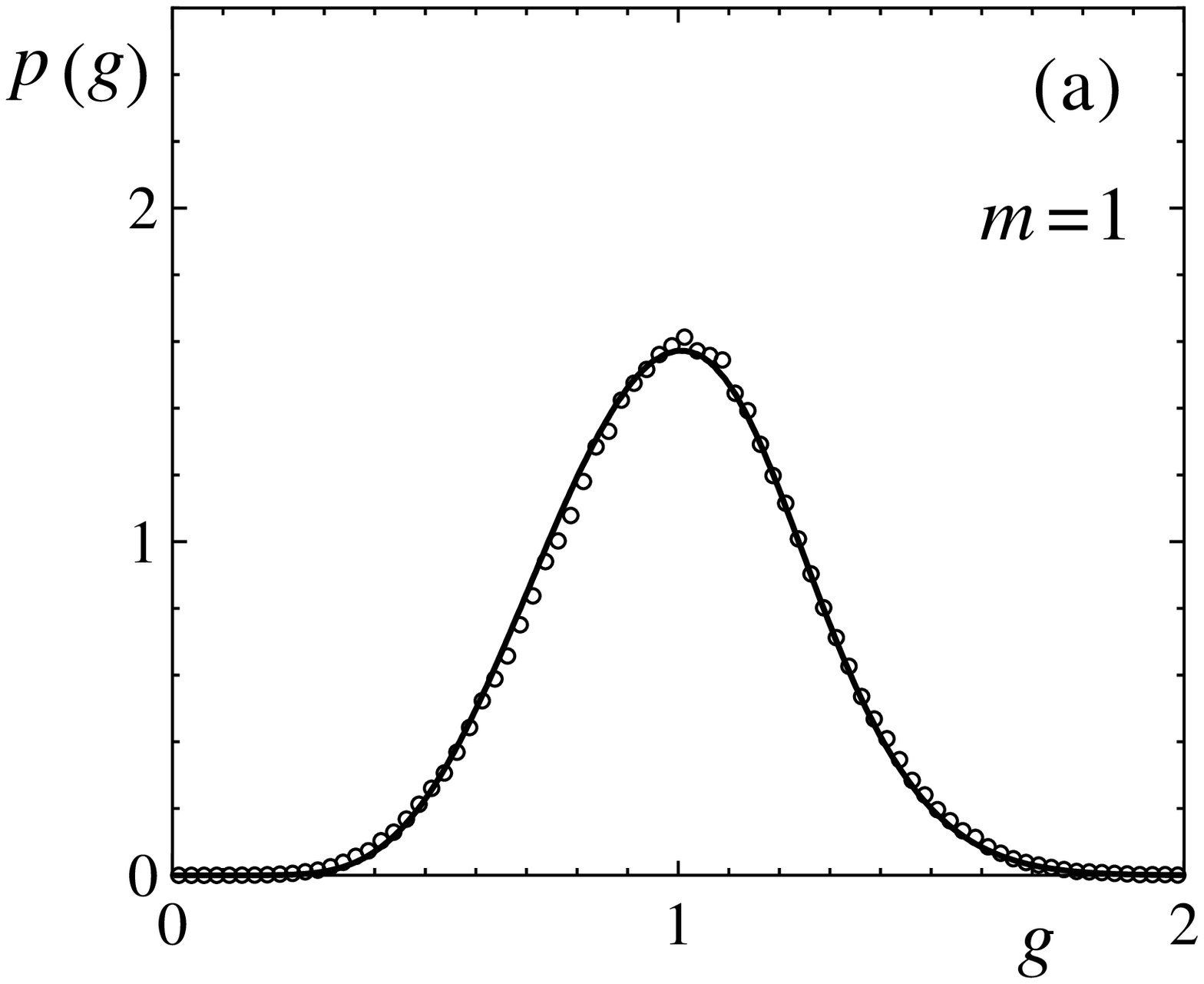}
\includegraphics[width=5.0cm]{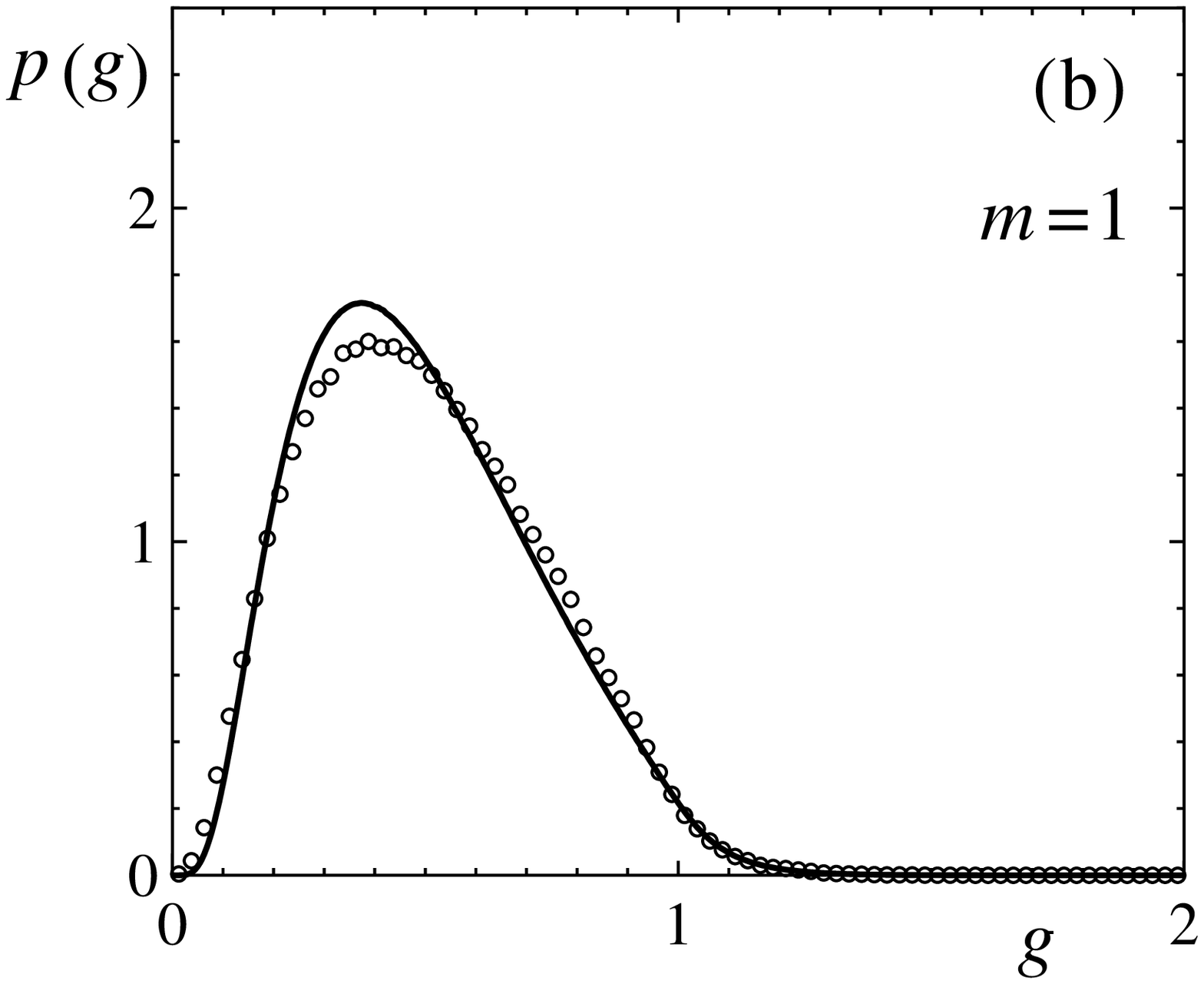}
\includegraphics[width=5.0cm]{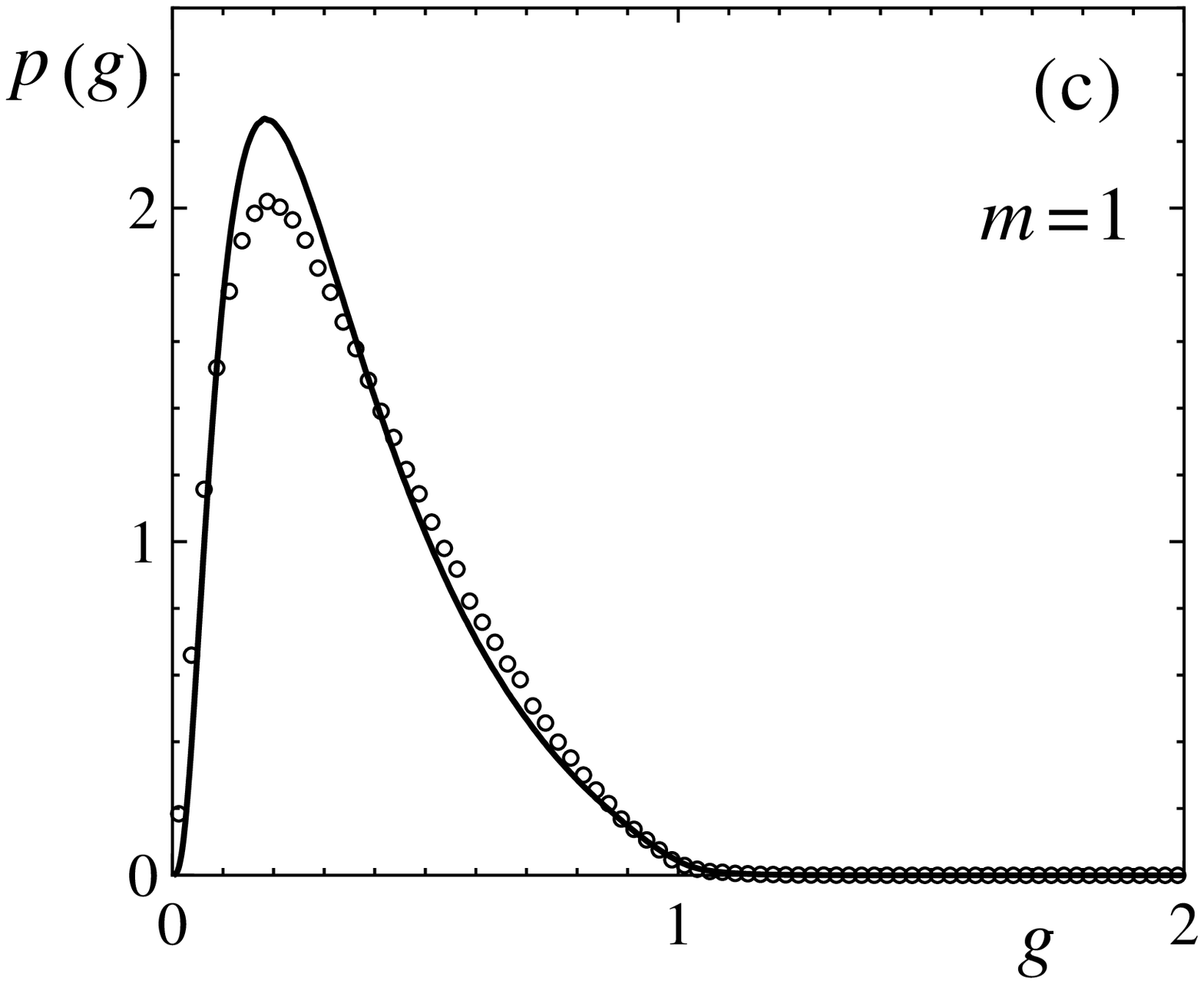}
\caption{
The conductance distributions for the case of $m = 1$
at (a) $\langle g \rangle = 1.0$, (b) $\langle g \rangle = 0.5$
and (c) $\langle g \rangle = 0.35$.
Open circles represent the numerical simulation result,
while solid lines represent the Monte Carlo result.
}
\end{center}
\end{figure}
We observe that the result of the numerical simulation approach
is qualitatively identical to that of the Monte Carlo approach
for both the cases of $m = 0$ and of $m = 1$.
Particularly, we again observe a kink near $g = 1$ in the case of $m = 0$
at $\langle g \rangle = 0.5$ and $0.35$,
while no such structure appears in the case of $m = 1$.

We finally note that a relatively large deviation between the two results
appears in the case of $m = 1$ for $\langle g \rangle = 0.5$ and $0.35$,
while no such deviation is observed in the case of $m = 0$.
This reflects the fact that the probability distribution $P(\{x_{a}\};s)$
given in eq.~(\ref{eq:prob_dis_mod}) overestimates
the influence of a perfectly conducting channel
in the long-wire regime of $s/N \gtrsim 1$~\cite{takane8}
although this is a very good approximation irrespective to $s/N$ in the case
of $m = 0$ without a perfectly conducting channel.~\cite{froufe-perez}

\section{Summary}

The dimensionless conductance $g$
in disordered quantum wires with unitary symmetry is studied for
both cases with and without a perfectly conducting channel.
To observe the influence of a perfectly conducting channel
on the behavior of $g$, we have calculated the conductance distribution $p(g)$
for these two cases in the crossover regime where
$\langle g \rangle$ is slightly smaller than unity.
We have adopted two approaches,
a classical Monte Carlo approach based on the existing scaling theory
and a numerical simulation approach with a tight-binding model.
In the latter approach, we have employed the tight-binding model for a graphene
nanoribbon with two zigzag edges (zigzag and bearded edges)
for the case with (without) a perfectly conducting channel.
We have confirmed that
the two approaches provide qualitatively identical results.
It is shown that in the absence of a perfectly conducting channel,
the distribution $p(g)$ are cut off when $g$ exceeds unity,
resulting in the appearance of a kink near $g = 1$, while no such structure
appears in the presence of a perfectly conducting channel.
This indicates that the absence of a kink in $p(g)$
is a notable characteristic of disordered quantum wires
with a perfectly conducting channel.

\section*{Acknowledgment}
K. W. acknowledges the financial support by a Grant-in-Aid for 
Young Scientists (B) (No. 19710082) and Specially Promoted Research
(No. 20001006) from the Ministry of Education,
Culture, Sports, Science and Technology, also by a Grand-in-Aid for
Scientific Research (B) from the Japan Society for the Promotion of Science
(No. 19310094).


\begin{thebibliography}{99}
\bibitem{puzzle} On the basis of a supersymmetric field theory,
Zirnbauer and co-workers~\cite{zirnbauer,mirlin} claimed that
the averaged dimensionless conductance $\langle g \rangle$
in the symplectic universality class converges to $1/2$
in the long-wire limit and thus Anderson localization is absent.
At present, we understand that the averaged conductance they obtained is
nothing but the arithmetic average of $\langle g_{\rm even} \rangle$
for the even-channel case and $\langle g_{\rm odd} \rangle$
for the odd-channel case.
See the discussion given in refs.~\citen{takane3} and \citen{sakai2}.
References~\citen{zirnbauer} and \citen{mirlin} should be regarded as
the first report that suggests a possibility of delocalization
in disordered wires with symplectic symmetry.
However, they recognized neither the even-odd difference nor
the presence of a perfectly conducting channel.
\end{thebibliography}
\end{document}